\documentclass[amsmath,amssymb,aps,pra,reprint,superscriptaddress,footnoteinbib,longbibliography]{revtex4-1}
\usepackage{amsmath,amssymb}
\usepackage[english]{babel}
\usepackage{ bbold }
\usepackage{upgreek}
\usepackage{dsfont}
\usepackage{graphicx}
\usepackage{comment}
\usepackage{braket}
\usepackage{colortbl}
\usepackage{blindtext}
\usepackage{xr-hyper}
\usepackage[hidelinks]{hyperref}
\usepackage{bm}
\allowdisplaybreaks

\begin{document}

\title{Quantum Optical Binding of Nanoscale Particles}

\author{Henning Rudolph}
\affiliation{University of Duisburg-Essen, Faculty of Physics, Lotharstra\ss e 1, 47057 Duisburg, Germany}

\author{Uro\v{s} Deli\'c}
\affiliation{University of Vienna, Faculty of Physics, Boltzmanngasse 5, A-1090 Vienna, Austria}

\author{Klaus Hornberger}
\affiliation{University of Duisburg-Essen, Faculty of Physics, Lotharstra\ss e 1, 47057 Duisburg, Germany}

\author{Benjamin A. Stickler}
\affiliation{Ulm University, Institute for Complex Quantum Systems  and Center for Integrated Quantum Science and Technology, Albert-Einstein-Allee 11, 89069 Ulm, Germany}

\begin{abstract}
Optical binding refers to the light-induced interaction between two or more objects illuminated by laser fields. The high tunability of the strength, sign, and reciprocity of this interaction  renders it highly attractive for controlling nanoscale mechanical motion. Here, we discuss the quantum theory of optical binding and identify unique signatures of this interaction in the quantum regime. We show that these signatures are observable in near-future experiments with levitated nanoparticles. In addition, we prove the impossibility of entanglement induced by far-field optical binding  in free space and identify strategies to circumvent this no-go theorem.
\end{abstract}

\maketitle

Optical fields provide unique opportunities for controlling the mechanical motion of dielectric objects in the quantum regime \cite{novotny2012}. Optical tweezers can stably suspend nanoscale to microscale particles \cite{Gieseler21,volpe2023roadmap}, where their center-of-mass and rotational dynamics can be monitored at the standard quantum limit by homodyning the scattered light \cite{millen2020a}. Trapped particles can be controlled by a combination of conservative dipole forces and nonconservative radiation pressure \cite{gonzalez2021}, and their rotations can be propelled via conservative and nonconservative optical torques \cite{stickler2021}. Combining the exquisite control with precise optical readout enables cooling the mechanical motion of submicron particles into the quantum regime with feedback techniques \cite{tebbenjohanns2021,magrini2021} or cavity cooling \cite{delic2020,ranfagni2022,piotrowski2023}.

If two or more dielectric particles are trapped close to each other they interact through the field scattered between them. Such {\it optical binding} interactions give rise to strong and tunable forces \cite{dholakia2010} that have been observed in numerous setups, ranging from atomic clouds \cite{maiwoger2022observation} and nanoscale particles \cite{arita2018,rieser2022,reisenbauer2023non,livska2023observations,vijayan2024cavity} to chains of microscale particles in fluids \cite{karasek2008long}. The dominant mechanism of optical binding is that the interference between the scattered light and the trapping light dynamically changes the optical forces acting on the particles. Importantly, this interaction can be strong, tunable, and nonreciprocal \cite{rieser2022}, meaning that they seemingly violate Newton's law of action equals reaction. This implies that the linearized particle dynamics close to the trap center are described by a non-Hermitian dynamical matrix, exhibiting exceptional points, unidirectional transport, and spontaneous symmetry breaking \cite{reisenbauer2023non}, with great potential for precision sensing \cite{metelmann2014quantum,mcdonald2020exponentially,budich2020non}.

\begin{figure}[b]
	\centering
	\includegraphics[width=1\linewidth]{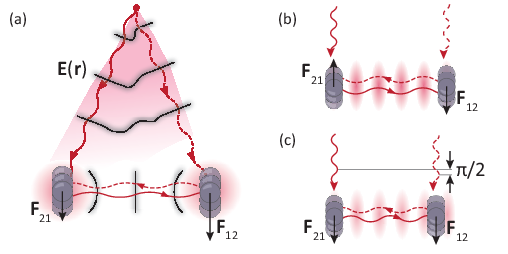}
    \caption{(a) Two nearby nanoparticles illuminated by a laser ${\bf E}({\bf r})$ interact through their scattered light fields. If the laser is phase coherent, the relative local phase can be used to tune the directionality and reciprocity of the mean interaction forces $\langle {\bf F}_{12}\rangle$ and $\langle {\bf F}_{21}\rangle$. In the quantum regime, this interaction leads to decoherence of the two-particle state and to shot-noise-induced correlations between the particles. (b) If the relative phase vanishes the mean interaction force is reciprocal $\langle {\bf F}_{12}\rangle  = -\langle {\bf F}_{21}\rangle $, while (c) the interaction becomes antireciprocal $\langle {\bf F}_{12}\rangle = \langle {\bf F}_{21}\rangle $ for the relative phase of $\pi/2$.}
	 \label{fig:drawing}
\end{figure}

To illustrate the richness of optical binding, let us consider the time-averaged force acting on a subwavelength sphere (dipole approximation) with polarizability $\alpha$ at position ${\bf r}_1$ in the laser field ${\bf E}({\bf r})e^{-ickt}$, with $k$ the wave number, due to the presence of an identical sphere at position ${\bf r}_2$ \cite{dholakia2010},
\begin{equation}\label{eq:optbindforce}
{\bf F}_{12} = \frac{\alpha^2}{2\varepsilon_0} \frac{\partial}{\partial {\bf r}_1} \text{Re}\left[ \mathbf{E}^*(\mathbf{r}_1)\cdot{\rm G}(\mathbf{r}_1-\mathbf{r}_2)\mathbf{E}(\mathbf{r}_{2}) \right].
\end{equation}
This force is nonreciprocal, ${\bf F}_{12} \neq -{\bf F}_{21}$, given that the free-space dipole Green tensor for $r>0$
\begin{equation*}
    {\rm G}(\mathbf{r}) = \frac{e^{ikr}}{4\pi}\left[ (1-ik r) \frac{3\mathbf{r}\otimes\mathbf{r}-r^2\mathbb{1}}{r^5}+ k^2 \frac{r^2\mathbb{1}-\mathbf{r}\otimes\mathbf{r}}{r^3} \right]
\end{equation*}
is complex-valued. The force being nonreciprocal, the resulting particle dynamics are nonconservative so that both momentum and energy are not conserved, but constantly supplied and extracted by light scattering (see Fig.~\ref{fig:drawing}). The recent progress in quantum cooling of levitated nanoparticles \cite{delic2019,tebbenjohanns2021,magrini2021,ranfagni2022,piotrowski2023} and the prospects of exploiting nanoparticle arrays in the deep quantum regime for superior force and torque sensing \cite{rudolph2022,rieser2022,carney2023searches,liska2023,reisenbauer2023non,vijayan2024cavity} raises the need for a full quantum mechanical description of optical binding interactions between co-trapped particles.

This article identifies unique quantum signatures of nonreciprocal optical-binding interactions and demonstrates how they can be observed in nanoparticle arrays. Specifically, we will see that optical binding in the quantum regime is described by a Markovian quantum master equation, which recovers the conservative and the nonconservative interaction between the nanoparticles and accounts for the ensuing decoherence. It allows us to rigorously derive three results of considerable experimental relevance: First, the quantum fluctuations of the light fields can induce pronounced correlations between the particles even when the particles would not interact classically. Second, non-Hermitian mode damping practically cannot cool the particle motion to the ground state. Third, free-space optical binding can never lead to entanglement between two co-trapped particles. We will discuss the empirical implications of these findings and present a concrete proposal to observe the quantum signatures of optical binding.

The quantum version of the optical binding interaction \eqref{eq:optbindforce} takes the form of a Markovian quantum master equation, which can be derived in five steps: (i) We determine the relation between the incoming light field ${\bf E}$ and the polarization field induced inside the two particles. This requires taking into account that the light scattered off one particle contributes to the polarization field inside the other particle. (ii) We calculate the optical forces acting on the two particles and uses them to derive the classical Lagrangian describing the combined dynamics of the two particles and the scattered fields. (iii) A mechanical gauge transformation renders the canonical momentum field independent of the particle degrees of freedom. (iv) One derives and canonically quantizes the classical Hamiltonian of the combined particle motion plus scattered fields via a Legendre transformation. (v) The scattered fields are traced out in the weak-coupling Born-Markov approximation. This procedure yields the Lindblad master equation for the two-particle state $\rho$,
\begin{align}\label{eq:master}
    \partial_t \rho =& -\frac{i}{\hbar}[H + V + V_{\rm opt},\rho]\nonumber\\ &+ \int d^2\mathbf{n}\sum_s \left( L_{\mathbf{n}s}\rho L_{\mathbf{n}s}^\dagger - \frac 1 2 \lbrace L_{\mathbf{n}s}^\dagger L_{\mathbf{n}s},\rho \rbrace \right).
\end{align}
Here, $H$ denotes the laser-independent Hamiltonian of the two particles, which may include additional trapping and manipulation fields, and $V$ is the laser-induced time-averaged dipole potential,
\begin{align}\label{opticalpotential}
    V = -\frac{\alpha}{4}\sum_{j=1}^2|\mathbf{E}(\mathbf{r}_j)|^2.
\end{align}
The optical binding interaction enters at two locations: The Lamb shift gives rise to the conservative optical binding potential
\begin{align}\label{conservativebinding}
    V_{\rm opt} =  -\frac{\alpha^2}{4\varepsilon_0} \sum_{j,j' =1\atop j\neq j'}^2\mathbf{E}^*(\mathbf{r}_{j'})\cdot\text{Re}[{\rm G}(\mathbf{r}_j-\mathbf{r}_{j'})]\mathbf{E}(\mathbf{r}_j).
\end{align}
It is determined by the real part of the Green tensor and can be interpreted as the interaction potential of two induced dipoles. This conservative interaction is always accompanied by the nonconservative optical binding interaction through the Lindblad operators
\begin{align}\label{lindbladians2}
    L_{\mathbf{n}s} =  \sqrt{\frac{k^3}{2\varepsilon_0 \hbar}}\frac{\alpha}{4\pi} \sum_{j=1}^2 \mathbf{t}_{\mathbf{n}s}^*\cdot\mathbf{E}(\mathbf{r}_j) e^{-ik\mathbf{n}\cdot\mathbf{r}_j}.
\end{align}
They are the coherent sum of the single-particle photon scattering amplitudes into direction ${\bf n}$ with orthogonal polarization directions $\mathbf{t}_{{\bf n}s}$, $s = 1,2$ \cite{rudolph2021}, reminiscent of superradiance \cite{lehmberg1970,lehmberg1970part2,agarwal1970,agarwal1971,gross1982,vogt1996}. The Lindblad dissipator describes nonconservative radiation pressure forces, decoherence, and the nonconservative optical binding interaction \cite{rudolph2021}. Together, the conservative and the nonconservative optical binding contributions in Eq.~\eqref{eq:master} recover the full optical binding interaction \eqref{eq:optbindforce}, i.e.\ $\partial_t \langle {\bf p}_1 \rangle = -\langle \partial V/\partial {\bf r}_1 \rangle + \langle {\bf F}_{12}\rangle$. More details on the derivation and its extension to an arbitrary number of  particles of arbitrary size, shape, and optical susceptibility can be found in Ref.~\cite{pra}.

\begin{figure}[b]
	\centering
	\includegraphics[width=0.8\linewidth]{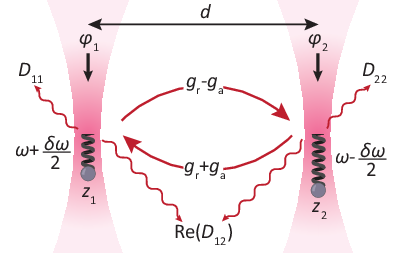}
    \caption{Two nanoparticles are harmonically trapped with frequencies $\omega\pm \delta\omega/2$ close to the foci of two optical tweezers driven by the same laser with phases $\varphi_{1}$ and $\varphi_{2}$. The light scattered off the particles (i) couples their motion nonreciprocally with coupling rates $g_{\rm r}+ g_{\rm a}$ and $g_{\rm r}- g_{\rm a}$, and (ii) imprints photon shot noise, leading to recoil heating with diffusion constants $D_{11}$ and $D_{22}$. The photon shot noise is correlated, as described by Re($D_{12}$).}
	 \label{fig:sketchplot}
\end{figure}

\begin{figure*}[t]
	\centering
	\includegraphics[width=1\linewidth]{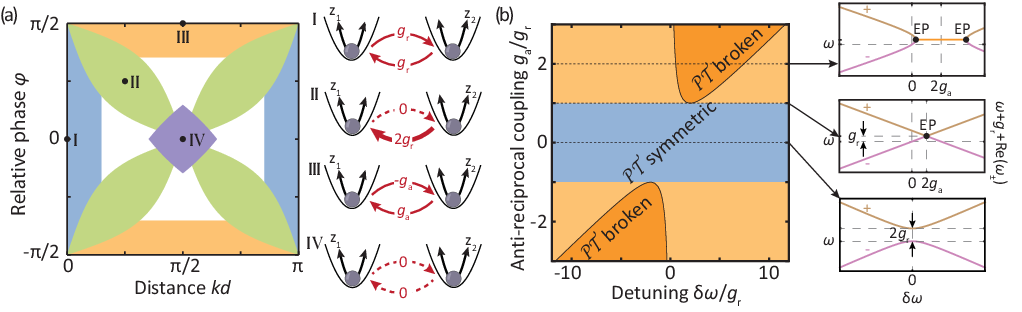}
	\caption{(a) Different regimes of quantum optical binding between two deeply trapped particles as a function of the relative tweezer phase $\varphi$ and the tweezer distance $kd$. The blue region indicates predominant reciprocal coupling $|g_{\rm r}|>|g_{\rm a}|$ and $|g_{\rm r}|>|\text{Re}(D_{12})|$; green indicates predominant directional coupling $|g_{\rm r} + g_{\rm a}|>2|g_{\rm r} - g_{\rm a}|$ or $|g_{\rm r} - g_{\rm a}|>2|g_{\rm r} + g_{\rm a}|$; orange indicates predominant antireciprocal coupling $|g_{\rm a}|>|g_{\rm r}|$ and $|g_{\rm a}|>|\text{Re}(D_{12})|$; violet indicates predominant recoil-noise correlations $|\text{Re}(D_{12})|>2\,\text{max}[|g_{\rm r} + g_{\rm a}|,|g_{\rm r}-g_{\rm a}|]$. The coupling corresponding to the points (I)-(IV) are depicted on the right. (b) Phases of broken and unbroken generalized ${\cal PT}$  symmetry as a function of the mechanical detuning $\delta\omega$ and the antireciprocal coupling $g_{\rm a}$. In the blue (orange) area the reciprocal coupling is greater (smaller) than the antireciprocal coupling, $|g_{\rm r}|>|g_{\rm a}|$ ($|g_{\rm a}|>|g_{\rm r}|$). ${\cal PT}$ symmetry is broken in the dark orange region, where the mode frequencies $\text{Re}(\omega_{\pm})$ are degenerate. The panels on the right show how the eigenfrequencies vary with mechanical detuning for different $g_a$ as indicated by the dotted lines. In the ${\cal PT}$-symmetry broken phase, two exceptional points (EPs) occur.}
	\label{fig:regimesplot}
\end{figure*}

For a physical intuition of quantum optical binding, we consider the case that both particles are deeply trapped in two separate optical tweezers at distance $d \gg 1/k$ (far-field approximation) and with relative optical phase $\varphi = \varphi_1 - \varphi_2$ (see Fig.~\ref{fig:sketchplot}). The motion along the tweezer axes can then be separated from the transverse motion because of a significant mismatch in mechanical frequencies. This simplifies the collective dynamics to two harmonic oscillators with dimensionless quadratures $z_{j} = \sqrt{m \omega} {\bf r}_j \cdot {\bf e}_z/\sqrt{\hbar}$ and $p_j = \sqrt{m} \dot{\bf r}_j \cdot {\bf e}_z/\sqrt{\hbar \omega}$, where $m$ is the particle mass and $\omega$ is the mean mechanical frequency so that $[z_j,p_{j'}] = i \delta_{jj'}$. Optical binding leads to coupling of the oscillators as characterized by the {\it reciprocal} and the {\it antireciprocal} coupling rates
\begin{subequations}\label{eq:couplings}
\begin{align}
    g_{\rm r} = & \frac{G}{k d} \cos( k d) \cos \varphi,\\
    g_{\rm a} = & \frac{G}{k d} \sin( k d) \sin \varphi.
\end{align}    
\end{subequations}
They decay as $1/k d$, as expected from far-field dipole radiation, and they oscillate both with the relative tweezer phase $\varphi$ and with the phase $k d$ as dictated by interference between tweezer and scattered light. The constant $G$ is determined by the tweezer power, tweezer polarization, and  particle polarizability \cite{rieser2022,reisenbauer2023non}, see also Appendix \ref{app:linear}. The two constants \eqref{eq:couplings} enter the linearized master equation together with interaction-independent oscillator dynamics $H_0$,
\begin{align}\label{eq:masterlin}
    \partial_t \rho = & -\frac{i}{\hbar}[H_0 + \delta H_0,\rho] + 2 i g_r [ z_1 z_2,\rho] \nonumber \\
    & + \sum_{j,j' = 1}^2 2D_{jj'} \left[ z_j \rho z_{j'} - \frac{1}{2} \{z_j z_{j'},\rho \} \right ].
\end{align}
Here, the second term in the first line accounts for the reciprocal interaction between the two particles. While the diagonal elements of the diffusion rate matrix $D_{11} \simeq D_{22}$ account for local shot noise heating, the off-diagonal element
\begin{align}\label{eq:diffrate}
    D_{12} = \frac{G}{k d} \sin (k d) \cos \varphi + i g_{\rm a} = D_{21}^*
\end{align}
gives rise to shot-noise correlations (real part) and antireciprocal coupling (imaginary part). Note that $D_{11}D_{22} > |D_{12}|^2$ so that the time evolution of $\rho$ is completely positive. The interaction also leads to a modification of the local trapping stiffness as described by
\begin{equation}
    \frac{\delta H_0}{\hbar} = (g_{\rm r} + g_{\rm a}) z_1^2 + (g_{\rm r} - g_{\rm a}) z_2^2.
\end{equation}
Figure \ref{fig:regimesplot} (a) demonstrates the tunability of optical binding, ranging from predominantly reciprocal coupling (blue), via unidirectional (green) and antireciprocal coupling (orange), to a regime dominated by shot-noise correlations (violet). The points indicate where the coupling becomes purely reciprocal [$g_{\rm a}=D_{12}=0$ at (I)], maximally unidirectional [$g_{\rm a}=g_{\rm r}$ at (II)], purely antireciprocal [$g_{\rm r}=\text{Re}(D_{12})=0$ at (III)], and where the recoil correlations are maximal [$g_{\rm r} = g_{\rm a} = 0$ at (IV)]. In the following, we will discuss three implications of this master equation for upcoming quantum optomechanical experiments with nanoparticles.

First, the master equation \eqref{eq:masterlin} predicts that the particles are subject to correlated shot noise as described by the real part of the diffusion rate \eqref{eq:diffrate}. This implies that the dynamics of the two particles becomes correlated even if both coupling rates \eqref{eq:couplings} vanish. This can be seen from the quantum Langevin equations for the mode operators in the co-rotating frame $a_j = (z_j + i p_j)e^{i(\omega +g_{\rm r})t}/\sqrt{2}$,
\begin{align}\label{eq:langevin}
    \frac{d}{dt} \left ( \begin{array}{c}
       a_1  \\
        a_2 
    \end{array}\right ) = & -i H_{\rm NH}  \left ( \begin{array}{c}
       a_1  \\
        a_2 
    \end{array}\right ) + \left ( \begin{array}{c}
       \xi_1  \\
        \xi_2 
    \end{array}\right )
\end{align}
where the dynamical matrix is
\begin{equation}\label{eq:hnh}
    H_{\rm NH} = \left ( \begin{array}{cc}
       -\frac{\delta \omega}{2} + g_{\rm a}  & -g_{\rm r} - g_{\rm a} \\
       - g_{\rm r} + g_{\rm a} & \frac{\delta \omega}{2} - g_{\rm a}
    \end{array}\right )
\end{equation}
with the mechanical detuning $\delta \omega$, which is controlled by the power difference of the tweezer traps, and we performed a rotating-wave approximation. The white-noise operators are correlated $\langle \xi^\dagger_{j'}(t') \xi_j(t)\rangle = 2 D_{jj'} \delta(t - t')$, implying that the particle motion becomes correlated even when $g_{\rm r} \simeq 0$ and $g_{\rm a}\simeq 0$, so that the dynamical matrix mediates no coupling. Intuitively, whether the two particles oscillate in phase (common mode) or in antiphase (differential mode) controls if their scattered fields interfere constructively or destructively, thus increasing or decreasing the recoil heating rates. As a consequence, the two normal modes thermalize to different temperatures in presence of external damping. For the particles from \cite{rieser2022}, the resulting difference between the normal modes' recoil heating rates is between 10-15\%. It could be observed by means of reheating measurements \cite{jain2016} if the noise on the particle is dominated by recoil heating, which occurs at pressures below $10^{-7}\,\text{mbar}$ in state-of-the-art experiments. It is a direct consequence of the correlation between the quantum noise forces (and thus cannot be obtained by naively adding single particle recoil heating to the classical equations of motion).

Second, non-Hermitian damping of the quasi-normal modes [right eigenvectors of \eqref{eq:hnh}] cannot prepare the particle motion in the ground state. For $g_{\rm a} \neq 0$, the dynamical matrix \eqref{eq:hnh} is non-Hermitian  with eigenfrequencies 
\begin{align}\label{eq:frequencies}
    \omega_{\pm} = & \pm \frac{1}{2} \sqrt{\delta \omega^2 + 4 g_{\rm r}^2 - 4 g_{\rm a} \delta \omega}.
\end{align}
For $|g_{\rm a}|<|g_{\rm r}|$ the normal mode frequencies are real and the dynamical matrix \eqref{eq:hnh} remains in the unbroken phase of a generalized ${\cal PT}$ symmetry \footnote{The generalized ${\cal PT}$ symmetry is given by the simultaneous swapping of the particles and the tweezers, so that $\delta \omega$ and $\varphi$ are replaced by $-\delta\omega$ and $-\varphi$ \cite{reisenbauer2023non}.}. However, the system can enter a symmetry-broken regime if $|g_{\rm a}| > |g_{\rm r}|$ when passing through an exceptional point at $\delta\omega = 2g_{\rm a} \pm2\sqrt{g_{\rm a}^2-g_{\rm r}^2}$. Then, the eigenfrequencies turn into a complex conjugated pair, so that one mode gives rise to an exponential growth in time, while the other leads to an exponential decay. The exponential suppression of mechanical motion is most pronounced for $g_{\rm r} = 0$ and $\delta\omega = 2 g_{\rm a}$. In this case, the Langevin equation \eqref{eq:langevin} describes the uncoupled dynamics of the two collective modes $(a_1\pm ia_2)/\sqrt{2}$ of the two oscillators at fixed relative phases $\pm \pi/2$. In the absence of quantum noise, the mode $(a_1- ia_2)/\sqrt{2}$ would decay exponentially, while $(a_1+ ia_2)/\sqrt{2}$ would increase exponentially. The unavoidable presence of quantum noise due to photon scattering acts as a finite temperature bath, forcing the decaying mode to saturate at the effective occupation $D_{11} kd/2G - 1/2 \geq 0$. In practice, the recoil rate clearly exceeds the coupling rate, $D_{11}\gg G/kd$, so that the stationary state of the decaying mode is far away from the ground state. Observing this saturation of the damped quasi-normal mode despite the absence of external heating sources, such as gas collisions or stray fields, presents a signature of quantum optical binding that will become relevant in near-future experiments. The phase diagram Fig.~\ref{fig:regimesplot} (b) demonstrates the occurrence of exceptional points and the time reversal symmetry-broken phase, as observed experimentally in Refs.~\cite{reisenbauer2023non,livska2023observations}.

Third, free-space optical binding cannot mediate entanglement. One could be tempted to expect that optical binding can induce entanglement between the two particles, given that changing the relative laser phase  with twice the mechanical frequency, $\varphi = 2\omega t$, transforms the coherent optical binding interaction in the co-rotating frame into the two-mode squeezing Hamiltonian
\begin{align}\label{eq:squeezing}
    H_{\rm r}(t) \simeq \frac{\hbar G}{2k d} \cos(kd) \left[a_-^2 + (a_-^\dagger)^2 \right].
\end{align}
Here $a_- = (a_2-a_1)/\sqrt{2}$. However, even if the antireciprocal coupling constant is tuned to zero, $g_{\rm a} = 0$, the optical-binding interaction can always be written as an LOCC (local operations and classical communication) channel by formulating a stochastic feedback quantum master equation, which models optical binding via a feed-forward loop of independent and local homodyne measurements and which yields Eq.~\eqref{eq:masterlin} in the ensemble average, see Appendix \ref{app:locc}. If such an LOCC description is possible the interaction cannot mediate entanglement, in agreement with findings for unidirectional quantum transport \cite{metelmann2017,clerk2022}. In free space, the vast majority of scattered photons contributes to decoherence while only a tiny fraction mediates the coherent optical binding interaction. However, if the effective recoil heating rate can be made smaller than the coupling rate, then optical binding would generate entanglement. Likewise, Eq.~\eqref{eq:squeezing} cannot squeeze below the ground state width if the coupling rate is smaller than the recoil heating rate, see Appendix \ref{app:locc}.
  
The conditions that preclude entanglement generation can be circumvented in several ways: (i) Recoil-scattering decoherence can be reduced by homodyning the position of both particles via detection of the backscattered light. The thus obtained measurement record may be used to determine the conditional quantum state of the system, which evolves according the optical binding master equation \eqref{eq:master} together with a stochastic measurement superoperator given in Appendix \ref{app:feedback}. If the fields are tuned to purely conservative optical-binding interaction, the effective recoil heating rate of the conditional state $D_{11}' = D_{11} (1 - \eta_{\rm det})$ is reduced according to the total detection efficiency $\eta_{\rm det}$. By achieving sufficiently large detection efficiencies, one might thus reduce the recoil heating rate below the conservative coupling rate $g_{\rm r}$ and thereby enable the generation of entanglement via optical binding. (ii) The local recoil rates in  Eq.~\eqref{eq:langevin} can be reduced by squeezing the electromagnetic vacuum state in the quadratures commuting with the local noise \cite{gonzalez2023}. In practice, this requires squeezing a single electromagnetic free-space mode for each particle, ideally with a large spatial overlap $\zeta$ with the scattered fields of the respective particles. The effective recoil heating rates are given by $D_{11}' = D_{11} [1-|\zeta|^2 (1-e^{-r})]$, where $r$ is the squeezing parameter. (iii) Finally, one may enhance the conservative optical-binding interaction by placing the particles in an optical cavity, realizing coherent scattering of both particles into the same cavity mode \cite{rudolph2020,vijayan2024cavity}. Since the cavity output can be detected with high efficiency, such a setup may be utilized for entanglement via single-photon detection and postselection \cite{rudolph2020}. Note that, for the particles from \cite{rieser2022}, roughly one out of ten photons contributes to optical binding, while the remaining light scattering leads to decoherence. In the future it might therefore be necessary to combine two or more of the mentioned techniques to enable entanglement via optical binding.

In summary, we presented the quantum theory of optical binding between two polarizable point particles and used it to discuss important implications for near-future experiments with levitated nanoparticles. The possibility for tunable two-mode squashing may well become relevant for measuring small interaction forces between the particles, e.g. due to gravity. We expect that the ability to continuously tune the interaction from fully reciprocal to fully nonreciprocal will render nanoparticle arrays an ideal platform for exploring and exploiting non-Hermitian quantum physics in near-future setups.

\begin{acknowledgements}
H. R., K. H. and B. A. S. acknowledge funding by the Deutsche Forschungsgemeinschaft (DFG, German Research Foundation)--439339706. B. A. S. acknowledges funding by the DFG--510794108 as well as by the Carl-Zeiss-Foundation through the project QPhoton. U. D. acknowledges support from the Austrian Science Fund (FWF, Project DOI 10.55776/I5111).
\end{acknowledgements}

\appendix

\section{Derivation of the linearized master equation}\label{app:linear}

In this section we sketch the derivation of the master equation \eqref{eq:masterlin}. We start by noting that the total laser field is the superposition of two optical tweezers of amplitudes ${\bf E}_{1,2}$,
\begin{align}
    \mathbf{E}(\mathbf{r}) = e^{ik{\bf e}_z\cdot{\bf r}} \left [\mathbf{E}_1   f_{\rm tw}(\mathbf{r}) + \mathbf{E}_2   f_{\rm tw}(\mathbf{r} - d {\bf e}_x)  \right ],
\end{align}
with identical tweezer field envelope
\begin{align}
    f_{\rm tw}(\mathbf{r}) = \frac{1}{1+i{\bf e}_z\cdot{\bf r}/z_{\rm R}} \exp\left( -\frac{({\bf e}_x\cdot{\bf r})^2 + ({\bf e}_y\cdot{\bf r})^2}{w^2(1+i{\bf e}_z\cdot{\bf r}/z_{\rm R})} \right),
\end{align}
whose foci are separated by the distance $d$ in the focal plane along ${\bf e}_x$. Since the Rayleigh range $z_{\rm R}$ is typically much greater than its corresponding beam waist $w$, the trapping frequencies for the motion in the focal plane are far detuned from the frequency along the optical axis. If the coordinates of the particles transverse to the beam propagation direction are deeply trapped, they can be omitted via the replacement $\mathbf{r}_{j} = \sqrt{\hbar/m\omega}z_{j} \mathbf{e}_z + d \delta_{j,2}$, with $z_j$ the dimensionless position operators for the motion along the optical axis. The linearized master equation \eqref{eq:masterlin} follows from the full quantum optical binding master \eqref{eq:master} by expanding the trapping field in the particle coordinates.

The kinetic energy of the particles can be written as $H = \hbar\omega \sum_j p_j^2/2$, with $p_j$ the dimensionless momentum operators for motion along the optical axis. Since both particles stay near the foci of their respective tweezers, and since the distance between the tweezers is much greater than the beam waist, the local tweezer dominates the field at each particle. It follows that the potential energy due to the laser beams can be harmonically expanded as
\begin{align}
    V \simeq -\frac{\alpha}{4}\sum_{j=1}^2 |\mathbf{E}_j|^2\left( 1 - \frac{\hbar z_j^2}{m\omega z_{\rm R}^2} \right),
\end{align}
from which the particle trapping frequencies follow as $m \omega_j^2 = \alpha |\mathbf{E}_j|^2/2 z_{\rm R}^2$.

In addition, the optical binding potential \eqref{conservativebinding} can be approximated around $z_j=0$ by taking the laser beam near the respective tweezer focus to be given predominantly by its plane wave contribution
$\mathbf{E}(\mathbf{r}_j)\simeq \mathbf{E}_j\exp[i(k-1/z_{\rm R}){\bf e}_z\cdot{\bf r}_j]$, where the Gouy phase reduces the local effective wave number by $1/z_{\rm R}$ \cite{gonzalez2019,rudolph2021}. Then, the optical binding potential is approximately (neglecting a constant offset)
\begin{widetext}
\begin{align}\label{eq:potexpand}
 V_{\rm opt} \simeq -\frac{\alpha^2 k^2 }{16 \varepsilon_0 \pi d} \cos(k d) \left(k-\frac{1}{z_{\rm R}}\right) \sum_{j,j' = 1 \atop j\neq j'}^2 \mathbf{E}_{j'}^*\cdot\left(\mathbb{1} - \mathbf{e}_x\otimes\mathbf{e}_x\right)\mathbf{E}_j \Biggl[ i\sqrt{\frac{\hbar}{m\omega}}(z_j-z_{j'}) - \left(k-\frac{1}{z_{\rm R}}\right)\frac{\hbar(z_j-z_{j'})^2}{2m\omega} \Biggr],
\end{align}
where we used that the particles interact predominantly via their scattered fields in the far field, so that we could neglect all contributions of order higher than $1/d$. The linear contribution to Eq.~\eqref{eq:potexpand} can be dropped because it only slightly shifts the trapping minimum. The quadratic term gives rise to reciprocal optical binding as well as to the a slight shift of the trapping frequencies. Assuming that both tweezers are polarized along the same direction with angle $\pi/2-\theta$ with respect to the tweezer connecting axis ${\bf e}_x$ yields
\begin{equation}
    V_{\rm opt} = \frac{\hbar\alpha^2  k^4|{\bf E}_1||{\bf E}_2|}{16\pi m \varepsilon_0 \omega d}\cos^2\theta \cos\varphi \cos(kd) \left ( z_1^2 + z_2^2 - 2  z_1 z_2 \right ),
\end{equation}
where we used that $k z_{\rm R} \gg 1$. Here, $\varphi$ denotes the relative tweezer phase.

The quadratic expansion of the Lindblad operators \eqref{lindbladians2} in the position operators $z_j$ reads
\begin{align}
    L_{\mathbf{n}s} \simeq & \sqrt{\frac{ k^3}{2\varepsilon_0\hbar}} \frac{\alpha}{4\pi}\sum_{j=1}^2 \mathbf{t}_{\mathbf{n}s}^*\cdot\mathbf{E}_j \Biggl[ 1 +i\sqrt{\frac{\hbar}{m\omega}}\left( k - \frac{1}{z_{\rm R}} - k \mathbf{e}_z\cdot\mathbf{n} \right)z_j -\left( k - \frac{1}{z_{\rm R}} - k\mathbf{e}_z\cdot\mathbf{n} \right)^2 \frac{\hbar z_j^2}{2m\omega} \Biggr].
\end{align}
\end{widetext}
Inserting this expression into the master equation \eqref{eq:master} yields the linearized master equation \eqref{eq:masterlin} after the following steps: first, we neglect the all terms cubic or higher order in $z_j$ as well as the linear shift of the potential minimum. Second, we neglect $1/z_{\rm R}$. Finally, we introduce
\begin{align}
    G = \frac{\alpha^2 k^5 |{\bf E}_1||{\bf E}_2|}{16 \pi m \varepsilon_0 \omega}\cos^2\theta,
\end{align}
yielding the master equation \eqref{eq:masterlin}.

\section{Free-space optical binding as LOCC interaction}\label{app:locc}

The linearized master equation \eqref{eq:masterlin} can always be understood as the ensemble average of a stochastic nonlinear quantum master equation, where the interaction is mediated via a feedback loop. In this description, the only coherent time evolution is due to the uncoupled oscillator Hamiltonian $H_0 + \delta H_0$. The interaction between the particles enters incoherently through two processes: (a) Their coupling to a common bath,
\begin{equation}
    {\cal D}\rho = \sum_{j,j'=1}^2 2D_{jj'}^{\rm ff} \left( z_j \rho z_{j'} - \frac 1 2 \lbrace z_{j'} z_j , \rho \rbrace \right),
\end{equation}
with the feed-forward diffusion matrix elements
\begin{subequations}\label{eq:diffmatrix}
\begin{align}
    D_{11}^{\rm ff} =& D_{11} - \Gamma - \frac{(g_{\rm r} + g_{\rm a})^2}{4\Gamma}
\end{align}
\begin{align}
    D_{22}^{\rm ff} =& D_{22} - \Gamma - \frac{(g_{\rm r}-g_{\rm a})^2}{4\Gamma},
\end{align}
\end{subequations}
\begin{widetext}
\noindent and $D_{12}^{\rm ff} = D_{21}^{\rm ff} = \text{Re}(D_{12})$. Here, $\Gamma$ determines the accuracy of homodyning the particle positions \cite{rudolph2022}, so that the stochastic measurement signals read $dy_j = \langle z_j \rangle dt + dW_j(t)/\sqrt{8\Gamma}$ with Wiener increments $dW_j(t)$. (b) The two particles are subject to a stochastic homogeneous feedback force described by the superoperator increment
\begin{equation}
    d{\cal F}\rho = - 2i\Big( (g_{\rm r} + g_{\rm a}) [z_1,\rho]\circ dy_2(t) + (g_{\rm r}-g_{\rm a}) [z_2,\rho]\circ dy_1(t) \Big)
\end{equation}
in Stratonovich calculus. Thus, the measurement signal $dy_j$ of particle $j$ determines the force on particle $j' \neq j$ and vice versa. Finally, the continuous measurement results in a stochastic localization of the particle state, as described by the superoperator increment \cite{rudolph2022}
\begin{align}\label{measurement}
    d{\cal C}\rho = -2\Gamma\sum_{j=1}^2 \Big(\lbrace z_j^2 - \langle z_j^2 \rangle, \rho\rbrace dt 
    & + 2\lbrace z_j - \langle z_j\rangle, \rho \rbrace \circ dy_j(t)\Big).
\end{align}
In total, the stochastic feedback quantum master equation for the  conditional state $\rho_{\rm c}$ reads
\begin{align}\label{feedforward}
    d\rho_{\rm c} = & -\frac{i}{\hbar}[H_0 + \delta H_0,\rho_{\rm c}]dt + {\cal D}\rho_{\rm c} dt + d{\cal F} \rho_{\rm c} +  d{\cal C} \rho_{\rm c}.
\end{align}
Converting this to It\^{o} form and taking the ensemble average $\rho = \mathds{E}[\rho_{\rm c}]$ yields the optical binding master equation \eqref{eq:masterlin}, proving that free-space optical binding can never mediate entanglement. Note that this equivalence between optical binding and a feed-forward loop requires that the measurement accuracy $\Gamma$ can be chosen such that the diffusion matrix $D_{jj'}^{\rm ff}$ is positive. For far-field-coupled particles, where $D_{jj}\gg |D_{12}|,|g_{\rm r}|,|g_{\rm a}|$, this is always possible. However, when the recoil heating rate falls below the conservative coupling rate, interpreting the interaction as a feed-forward loop would require effective negative decoherence rates, see \eqref{eq:diffmatrix}, violating the complete positivity of the time evolution.

In addition, note that the impossibility of creating entanglement in free space implies the impossibility of generating a two-mode squeezed state \footnote{We define squeezing as a reduction of the fluctuations of one mechanical quadrature below the zero-point uncertainty of $1/2$.} via the squeezing Hamiltonian \eqref{eq:squeezing} if the recoil noise in Eq.~\eqref{eq:masterlin} exceeds the coupling strength. This can be seen by deriving the Ehrenfest equations of motion of Eq.~\eqref{eq:masterlin} for the first and second moments of the relative position  $z_-=z_2-z_1$ and momentum operators $p_- = p_2-p_1$ for the squeezing setup discussed in the derivation of Eq.~\eqref{eq:squeezing}. They imply that the variances of the two canonically conjugated quadratures $Z_\pm =(z_- \pm p_-)/\sqrt{2}$ follow
\begin{align}
    \partial_t \text{var}(Z_\pm) = D_{11} \mp 2\frac{G}{kd}  \text{var}(Z_\pm),
\end{align}
provided we choose conservative coupling with $\cos(kd) = 1$. It shows that the difference mode is squashed in one quadrature while its canonically conjugate is amplified. Eventually, the squashed quadrature will reach the stationary value $D_{11}kd/2G$, which exceeds $1/2$ since the free-space recoil rate is greater than the coupling rate. A two-mode squeezed state can still be prepared by decreasing the recoil rate or increasing the coupling along the lines of the discussion in the main text concerning the generation of inter-particle entanglement.

\section{Recoil noise reduction via homodyne detection}\label{app:feedback}

Here we show how continuous homodyning effectively reduces the recoil heating rate, as stated in the main text. We start by considering a single particle,  the generalization to two particles is straightforward and will be discussed afterwards.

Homodyning the scattered field with net detection efficiency of $\eta_{\rm det}$ and local-oscillator phase $\phi$ measures the particle position $z$ with signal $dy(t) = \langle z \rangle\cos\phi \,dt + dW(t)/\sqrt{8 D \eta_{\rm det}}$, where $D$ is the recoil diffusion constant and $dW(t)$ is a Wiener increment. The measurement backaction enters the dynamics of the conditional state $\rho_{\rm c}$ through a stochastic term in the quantum master equation in It\^{o} form \cite{rudolph2022}
\begin{align}\label{stochasticmaster}
    d\rho_{\rm c} = \mathcal{L}_{\rm opt}\rho_{\rm c} dt + \sqrt{2D\eta_{\rm det}} \left( e^{i\phi}z\rho_{\rm c} + e^{-i\phi}\rho_{\rm c} z - 2\cos\phi\langle z \rangle \rho_{\rm c} \right) dW(t).
\end{align}
Here, $\mathcal{L}_{\rm opt}\rho$ is the right-hand side of the optical binding master equation \eqref{eq:master}. For an infinitesimal time step $dt$, this master equation is solved by
\begin{equation}\label{timeevolution}
\rho_{\rm c}(t+dt)= \frac{1}{{\cal N}} \int_{-\infty}^\infty du \exp \left ( -\frac{u^2}{2} \right ) \mathcal{W}(u)\rho_{\rm c}(t)\mathcal{W}^\dagger(u)    
\end{equation}
where ${\cal N}$ is a normalization constant and the operators \cite{rudolph2022}
\begin{align}\label{krausoperator}
    \mathcal{W}(u) = \exp\left[ -\frac{i}{\hbar}H_{\rm det} dt  -2D\eta_{\rm det} (z\cos\phi - dy(t)/dt )^2 dt - iu z \sqrt{2D(1-\eta_{\rm det})dt}\right].
\end{align}
The state-dependent and stochastic hermitian detection operator is given by
\begin{align}\label{hdet}
    H_{\rm det} = H_0 + 2\hbar D\eta_{\rm det}\cos\phi\sin\phi\, z^2 - \hbar\left(\sqrt{2D\eta_{\rm det}} \sin\phi \frac{dW(t)}{dt}  + \frac{4D\eta_{\rm det}}{\hbar}\cos\phi\sin\phi \, \langle z \rangle \right)z,
\end{align}
implying that gaussian states remain gaussian given that $H_0$ is quadratic \cite{rudolph2022}. 
\end{widetext}
The operator \eqref{hdet} accounts for $\eta_{\rm det} \neq 0$ for the coherent impact of the measurement process. It is reversible if the measurement record $\langle z\rangle$ is available, and it reduces to the optical-binding dynamics for $\eta_{\rm det} = 0$. In contrast, the second term in Eq.~\eqref{krausoperator}, which is Gaussian in $z$, describes spatial localization of the state due to the measurement process. The final term in Eq.~\eqref{krausoperator} describes a homogeneous force for fixed $u$ and thus accounts for recoil heating with the effective diffusion constant $D(1 - \eta_{\rm det})$ after the average over $u$. For $\phi=\pm\pi/2$ the measurement-induced localization vanishes, as can be seen  by noting that \eqref{krausoperator} becomes unitary, so that the spatial localization of the particle is determined only by recoil heating with effective diffusion constant $D(1 - \eta_{\rm det})$. (Rewriting the feedback master equation \eqref{stochasticmaster} in Stratoniovich form shows that also for general $\phi$ recoil heating is determined by this effective diffusion constant.) 

The single-particle description can be generalized straightforwardly to the detection of two particles interacting via purely conservative optical binding, $D_{12}=0$, and with diffusion constants $D_{11}$ and $D_{22}$. For this, one adds a second stochastic term to Eq.~\eqref{stochasticmaster}, which describes the detection of the second particle position with independent measurement noise. The resulting effective diffusion coefficients given the detection efficiency $\eta_{\rm det}$ are $D_{11}(1 - \eta_{\rm det})$ and $D_{22}(1 - \eta_{\rm det})$, respectively. Note that measuring the two particles independently with high detection efficiencies ($\eta_{\rm det}$ near 1) may require resolving the angular distribution of the scattered light. This is because the scattering field of the two particle overlap, limiting the achievable degree of confocal mode matching. Therefore, the homodyne detection measures in general a linear combination of $z_1$ and $z_2$, as determined by the overlap of the scattered fields with the local oscillator. The corresponding master equation can be derived through a straight-forward generalization of the above argument, as discussed in Ref.~\cite{rudolph2022}.

\bibliographystyle{myapsrev}

\end{document}